\catcode`\=13
\def{\relax}

\newcommand{\etal}  {{\em et al.}}

\documentclass[]{spie}  

\usepackage[]{graphicx}

\title{Mass-producing spectra: The SDSS spectrographic system}

\author{Peter~R.~Newman\supit{a}, %
Dan~C.~Long\supit{a}, %
Stephanie~A.~Snedden\supit{a}, %
S.J.~Kleinman\supit{a}, %
Atsuko~Nitta\supit{a}, %
Michael~Harvanek\supit{a}, %
Jurek~Krzesinski\supit{a,b}, %
Howard~J.~Brewington\supit{a}, %
J.C.~Barentine\supit{a}, %
Eric~H.~Neilsen, Jr.\supit{a,c} and %
David~J.~Schlegel\supit{d}, %
for the SDSS Collaboration %
\skiplinehalf %
\supit{a}SDSS Observer, Apache Point Observatory, P.O. Box 59, Sunspot, NM 88349, U.S.A.; \\
\supit{b}Mt. Suhora Obs., Cracow Pedagogical Univ., ul.
Podchorazych 2, 30-084 Cracow, Poland; \\
\supit{c}Fermi National Accelerator Laboratory, P.O. Box 500, Batavia, IL 60510, U.S.A.; \\
\supit{d}Department of Astrophysical Sciences, Princeton University,
Princeton, NJ 08544, U.S.A.
 }

\authorinfo{Send correspondence to P.R.N.: {\tt prn@apo.nmsu.edu}, {\tt www.apo.nmsu.edu}, +1(505)437-6822}

\begin{document}
  \maketitle

\begin{abstract}
The Sloan Digital Sky Survey is the largest redshift survey conducted to
date, and the principal survey observations have all been conducted on the
dedicated SDSS 2.5m and 0.5m telescopes at Apache Point Observatory. While
the whole survey has many unique features, this article concentrates on a
description of the systems surrounding the dual fibre-input spectrographs
that obtain all the survey spectra and that are capable of recording 5,760
individual spectra per night on an industrial, consistent, mass-production
basis.  It is hoped that the successes and lessons learned will prove
instructive for future large spectrographic surveys.
\end{abstract}



\section{The start of the Industrial Revolution}
\label{sec:intro}

The industrial revolution in astronomical spectroscopy, as with many
revolutions, has taken some time to get up to speed.  It can be traced
back to the seminal work by Hill and collaborators\cite{hilletal80} on
simultaneous spectroscopy of galaxies using the Medusa spectrograph with
32 targeted optical fibres on the 2.3m telescope at the Steward
Observatory in the late 1970's and early 1980's, with which they succeeded
in gathering useful spectra over wavelengths of $\sim3900$--$5500$~\AA\ in
3-hour exposures. By the early 1990's, the hand-crafted instrument of
Steve Shectman and collaborators was at the forefront of the revolution,
"strip-mining" the sky with the 128-fibre {\em Fruit and Fiber\/}
spectrograph on a 2.5m telescope and celebrating obtaining several hundred
spectra per night during the Las Campanas Redshift Survey.
\cite{shectmanetal92,shectman93,shectmanetal96} This was an order of
magnitude improvement in observing efficiency in about a decade.

To date, some three dozen multi-object fibre spectrographs and variants
have been built or proposed and the multiplex advantage of simultaneously
reaping redshifts from large numbers of galaxies in each telescope field
of view has remained a powerful force driving forward instrument system
design.  While many of the ideas introduced by Hill \etal\ -- e.g.\
drilled-plate fibre mounting, cartridge interchange, coherent bundles of
fibres for imaging guide stars -- remain familiar in one form or another
in many of these instruments, developments in robotics have permitted
several other classes of spectrograph design to appear (see e.g.\ Refs.
\citenum{ingerson93} and \citenum{lewisetal2002}).

Improvements in efficiency have continued so that, today, we can obtain
spectra of galactic and extragalactic objects nearly an order of magnitude
faster still than in the 1990's, over a wider wavelength range and with
higher spectral resolution. Although the Sloan Digital Sky Survey (SDSS)
is using a 2.5m telescope, of similar size to that used by Shectman and
collaborators, we have a much larger array of fibres (640 against 128),
larger and more efficient detectors ($4 \times 2048 \times 2048$ pixels on
integrating CCDs against $1 \times 2048 \times 1520$ pixels on a photon
counter) that have reduced exposure times by half, and much less overhead
in the observing procedures.  The result is that we can observe up to 9
target fields per night now versus 4-5 then to similar limiting
magnitudes, and gather 5 times more targets per field, covering
wavelengths of $\sim3800$--$9100$~\AA.  We have also learned how to
routinely achieve a spectrophotometric calibration for fibre spectra of a
few percent.\cite{abazajianetal2004}

Indeed, on each of the nights of MJD 52669 (2003 January 29/30) and MJD
53084 (2004 March 19/20), the SDSS observers at Apache Point Observatory
(APO) obtained 5,760 spectra using the SDSS systems -– possibly the most
individual spectra ever observed per night\footnote{If anyone is aware of
a larger count, please contact P.R.N.} so far.  But there may be more to
come: radical design concepts for future fibre-input spectrographs on
telescopes of up to 8-m aperture, such as LAMOST\cite{suc2003} and
KAOS,\cite{deyetal2003} offer the possibility of further increasing
productivity by perhaps another order of magnitude in the next decade.

Details of the SDSS telescope, instruments and observing procedures have
been well documented elsewhere.\cite{gunnetal98,yorketal2000,long2002}
This article focuses on the elements that were critical to the success of
the SDSS in obtaining the spectra at the best rate: efficient equipment
and observing procedures; flexible observing plans; real-time quality
assurance, and, most important, a highly skilled and motivated team of
people.

\section{Efficient observatory hardware}
\label{sec:hardware}


\begin{figure}
  \centering
  \includegraphics[clip,width=6.75in]{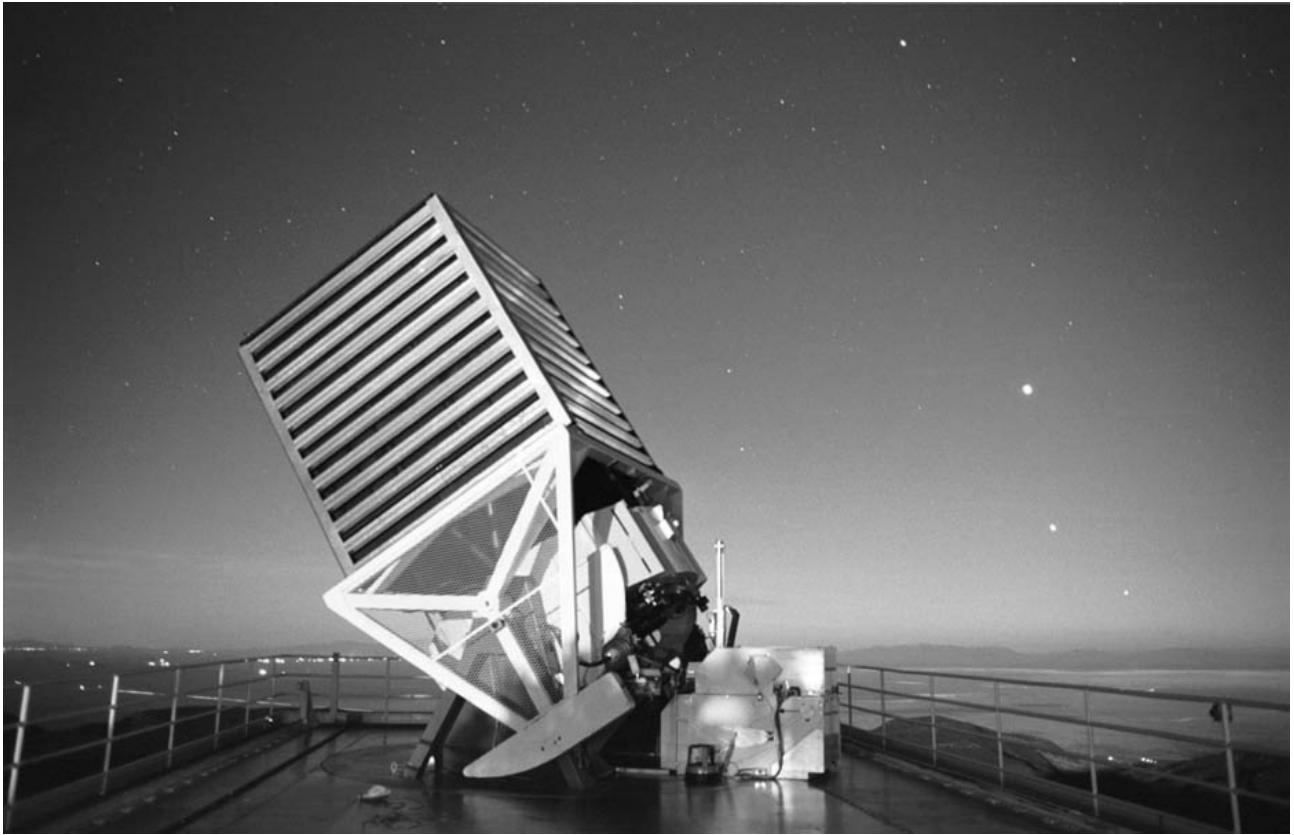}
  \caption{The SDSS 2.5m telescope, showing the slatted wind baffle
  enclosure and the imager's storage``doghouse''.
  Photograph copyright \copyright\ 1999 Dan Long. }
  \label{fig:telescope}
\end{figure}

\subsection{Enclosures and thermal management}

The SDSS 2.5m telescope has two separate enclosures that divide the
functions of a conventional ``dome'' type of enclosure.  The larger
secondary enclosure, whose principle functions are to protect the
telescope from weather and facilitate engineering operations, consists of
a run-off building (``shed'') which is moved on rails about 25~m from the
telescope prior to nighttime operations. Removing the enclosure and
refilling instrument-mounted liquid nitrogen dewars (necessary because the
telescope dips to 6~deg altitude in the process) is relatively slow,
taking $\sim 20$~min, but has little effect on observing efficiency unless
weather conditions are changing rapidly.

This enclosure includes a powerful air-conditioning system which is
actively managed with the objective of having the telescope and, in
particular, the mirrors in thermal equilibrium with the expected outside
air temperatures at twilight.  To this end, both the primary and secondary
mirrors carry a large array of thermometers, and much effort has been
expended on improving the forced ventilation of the primary mirror to the
point where we can easily track normal changes in air temperature with
mirror-to-air temperature differences and cross-mirror gradients on the
primary rarely exceeding 0.5~K during operations.  The secondary mirror
was found to be prone to super-cooling under clear sky and now carries a
thermal shield on its rear (sky-facing) surface, so that the front-to-back
temperature gradient is normally $< 0.2$~K.  Mirror temperatures (along
with many other telescope hardware measurements) are archived and
displayed in real-time to the observers with a sophisticated Telescope
Performance Monitor system.\cite{mcgeheeetal2002}

The much smaller primary enclosure, known as the wind baffle, protects the
telescope from wind-shake and stray light (see fig.\ \ref{fig:telescope}).
This is unusual in that it is ``shrink-wrapped'' onto the telescope, with
an average separation from the telescope of a just few centimetres.  The
wind baffle is moved by a separate servo system slaved to the telescope
position. This provides advantages in that the enclosure matches the slew
rate of the telescope (2~deg sec$^{-1}$ in altitude and azimuth) and
provides consistent and stable protection from buffeting in wind speeds in
excess of 55~kph.  The wind baffle also carries moveable flat-field screen
"petals" atop the secondary mirror truss that can move into and out of the
field of view in $\sim 5$~sec. These permit obtaining rapid in-situ
flat-field and arc lamp calibration exposures as required.

\subsection{Optical design}

The SDSS 2.5-m telescope has an f/5 modified Ritchey-Chr\'{e}tien optical
design, albeit with a wide (3-deg) field of view,\cite{yorketal2000} but
it has at least one unusual feature in that the primary mirror can be
pistoned a significant distance along the optical axis relative to the
instrument mount to provide an image scale adjustable by $\pm0.05$ per
cent.  This allows the observers to simply compensate for any mismatch
between the temperature for which each aluminium plug-plate was designed
and the actual temperature at the time of observation.  Determination of
the required scale is discussed in \S\ref{sec:guiding} below.

\subsection{Optical fibre mount plates}

The SDSS spectrographs use nine identical cartridges each carrying 640
science fibres and 11 coherent fibre bundles for taking images of guide
stars.\cite{owenetal98}  These cartridges allow rapid change of the
spectrographic fields targeted by the telescope.  One end of each fibre is
plugged into a drilled aluminium plate that is placed at the telescope
focal surface. The other end of the science fibres terminate on a pair of
pseudo-slits, each of which carry 320 fibres into the entrance apertures
of the dual spectrographs.  The coherent fibre bundles terminate on a
block that is imaged by a separate camera used for guiding. The plug
plates are drilled\cite{siegmundetal98} with the positions of targets
selected from imaging data.  The imager\cite{gunnetal98} is thus an
integral part of the spectrographic system, having the same field of view,
detecting objects to $\sim 2$ magnitudes fainter than the spectrographic
survey limits.  The area covered by the plates will eventually tile the
full survey area.\cite{blantonetal2003}

Plates typically carry $\sim 590$ fibres on survey targets plus $\sim 50$
fibres on calibration stars and blank sky, plus the 11 guide fibre
bundles.  Each plate takes $\sim 70$~min to observe (depending on sky
conditions) including field acquisition, flat-field and arc calibration
exposures, typically $3 \times 15$~min pointed target exposures and
changing the cartridge mounted on the telescope.

The SDSS expects to use $\sim 2000$ plates, so changing plates needs to be
a consistent and efficient operation.  The day after each night of
spectrographic observations, the APO fibre-plugging crew, working in a
support building at the end of the telescope pad, removes all fibres from
the plates for which observations are complete, puts a new plate on the
cartridge, inserts the fibres into the new plate, adjusts the plate
profile to match the focal surface curvature and maps the correspondence
between fibre and slit positions.  The cartridge is then ready for more
observations the next night.  Adjusting the plate profile is particularly
important for efficient operations, as variation in focus and fibre
telecentric angle across a plate can have large effects on the rate at
which signal is accumulated.\cite{newman2002}

A special plugging station lifts and inverts the cartridge, allowing two
people to work comfortably together inserting fibres downwards into the
new plates, with target fibres plugged into any hole then automatically
mapped from slit-head position to plate position.  This mapping is
achieved by back-illuminating the fibres with a laser one at a time from
the slit head and computer-processing images of the plug plate from a
digital camera to identify which fibre lights up.  All nine plates, 5,760
object fibres and 99 guide fibres can be replaced in $\sim 6$ hours during
the day ready for the next night's observations.

\subsection{Instrument change}

Instrument change, both from one spectrograph cartridge to the next and
between spectroscopy and imaging, is naturally a frequent activity during
SDSS night-time operations, so much careful thought has gone into making
the process both fast and foolproof.  The telescope has an
altitude-azimuth mount with instruments necessarily carried on a driven
rotator. All equipment changes are performed with the telescope pinned
pointing at the zenith, and equipment is lowered and lifted with a
hydraulic lift permanently mounted in the cone, so only the equipment
actually being changed has to be moved to and from the telescope.

Switching spectrograph cartridges is a one-person operation.  It involves
simply unlatching and lowering one 150-kg cartridge with the hydraulic
lift onto a dual-cartridge cart, moving the cart so the next cartridge is
under the mount point, then lifting and latching the new cartridge.  The
typical time between issuing the command to slew to the instrument change
position when exposures for one cartridge are completed and issuing the
command to slew to the target field for the next cartridge is under three
minutes, and rarely is it more than four minutes.  Most of this time is in
the slew.  Standard procedure is presently to complete read-out of the
last exposure for a field before starting the slew to instrument change,
as there was concern about increasing electrical noise, but we have been
experimenting with starting the slew as soon as read-out starts, without
any apparent increase in noise in the data. This parallelism saves $\sim
2$~min per plate.  The cartridge removed is then returned to the support
building for plate replacement the following day.

Changing mode from spectroscopy to imaging is somewhat more complex.  The
imager\cite{gunnetal98} is heavy, $550$--$570$~kg depending on the
contents of the secondary liquid nitrogen dewars, and is connected to the
rest of the world by an umbilicus $\sim 10$~cm in diameter.  Moving both
of these safely make mounting and dismounting the imager a two-person
operation.  Again with the telescope pinned at zenith, the spectrograph
cartridge is removed, then a second cart is lifted to carry away the
spectrographic corrector lens (leaving a common corrector in place).  When
not mounted on the telescope, the imager lives in its ``doghouse'' that is
mounted on the rotating floor on the opposite side from which
spectrographic equipment is loaded (see fig.\ \ref{fig:telescope}).  Once
the spectrograph cartridge is removed, the doghouse is opened and the
imager cart is pulled out on rails onto the hydraulic lift which lifts it
into place.  The imager and spectrographic cartridges share a common set
of latches, although the imager also has addition latches for the saddle
that carries its liquid nitrogen supply and ancillary electronics.

Instrument change clearly involves moving and mating heavy and expensive
equipment.  As with all operations involving humans, especially when
working against the clock, usually in the dark, and often in cold weather,
any risk of damage to operator or equipment due to human error is to be
avoided.  For this reason, all telescope and instrument motions are
heavily interlocked to, as far as is possible, prevent motions in the
wrong sequence.  For example, the interlocks sound an alarm if the
doghouse is opened when the T-bars supporting the imager CCDs are not in
the ``transport'' position, and they prevent motion of the lift to raise
the imager unless the spectrograph corrector lens has been removed.
Although it can be troublesome for the observers when interlocks
unexpectedly prevent motion, they do improve confidence when the observers
are trying to minimise the time taken to change instruments to maintain
observing efficiency.

\subsection{Support instruments for spectroscopy}

There are two ancillary instruments at APO that are run in support of
efficient spectrographic operations.\footnote{Live data from both
instruments described here are available nightly on the APO web site, {\tt
www.apo.nmsu.edu}.  No discussion of the SDSS 0.5m photometric telescope
is included here, as its function is to support imaging operations and is
hence outside the scope of this article.} First, we run a differential
image-motion monitor\cite{restetal2000} (DIMM) to quantitatively estimate
the true seeing independent of telescope-induced degradation of the
point-spread function. This lets us make sensible decisions as to the
thermal state of the mirrors and in choosing between imaging and
spectroscopy.  The DIMM is mounted on the end of the 2.5-m telescope pier,
so it samples the actual seeing close to the main telescope.

Secondly, an infrared sky camera provides still images of whole sky every
30~sec from which rolling 30-minute movies are constructed, both of which
have the current pointing of the telescope superimposed. This provides the
observers with a detailed view of current and recent sky conditions, and
quantitative measurements of the transparency of the sky based on the
distribution of the histogram of pixel values.  We use this both for
general decisions about which mode of observing is most appropriate, and,
particularly when imaging, to see if any isolated clouds have passed over
the field of view of the telescope and to decide if the variance in
transparency across the sky is small enough to declare conditions
photometric.

\section{Efficient observing systems and procedures}
\label{sec:systems}

\subsection{Observing staff}

There are presently nine staff at APO who rotate through SDSS observing
shifts, although three have other duties that limit their night-shift
time: one (S.J.K.) is largely involved in operations planning and
coordination, another (D.C.L.) has responsibilities for overall quality
assurance on the mountain, and a third (J.C.B.) is also an Observing
Specialist on the ARC 3.5m and NMSU 1m telescopes.  Our normal mode of
operation is for rolling pairs of observers to staff each night shift,
each person working for 2--6 nights in a row.  Since early 2003, we have
also operated a day shift, staffed by one of the observers in monthly
rotation.

\subsection{Preparation before nightfall}

Efficient night-time operations\cite{long2002} begin with a working
telescope and instruments.  The observer on day shift is responsible for
ensuring all system are operational before sunset in liaison with the APO
engineers. Pre-flight checks include full operational tests of the
telescopes, instruments and data acquisition systems, plus mounting the
first instrument for the night and fine adjustment of the spectrographs'
collimation and focus.

Although on the vast majority of nights, all systems are ``go'', catching
the rare problems that do come up during daylight is clearly advantageous
for efficiency. In addition, stress is reduced when the night shift can be
confident of being greeted with a working system (or at least one with
known problems!).

Observing science targets takes place while the sun is $\sim 12$~deg or
more below the horizon.  When we start the night with spectroscopy, we
will usually acquire the first field before the end twilight, so we can
make the initial focus and image scale adjustments before dark.

\subsection{Target acquisition, guiding and exposure control}
\label{sec:guiding}

Provided the pointing model and associated encoder scales are updated
regularly to compensate for thermal variations, the SDSS telescope is
capable of placing guide stars into (or very close to) the coherent guide
fibres for any plate on command.  We normally update the pointing model at
the start of each dark run.  The observers do not have to deal with target
positions in terms of sky coordinates.  Instead, the control software is
told which cartridge and plate are mounted, and the sky coordinates are
automatically loaded from the plate database.  This avoids all
transcription errors and makes target acquisition very straightforward.

Once two or more guide stars are visible in the display showing the images
from the coherent fibre bundles, an autoguider takes control of all three
telescope axes.  This will centre the guide stars and maintain pointing to
within $\sim 0.1$~arcsec with $\sim 15$~sec resolution.  The autoguider
also calculates and reports the required change in telescope image scale
to match the plug plate temperature.  This is not part of the automated
loop, as there is significant image motion when the primary mirror is
pistoned to adjust the scale.  However, once the observer has set the
image scale, it normally remains within the permitted
tolerance\cite{newman2002} of $\pm 0.003$ per cent for the duration for
exposures for a plate.  The mirror servos automatically maintain focus
during scale changes, so provided the telescope remains in thermal
equilibrium, no further focus adjustment is needed from the observers when
switching between plates designed for different observational
temperatures.

The autoguider also reports the apparent seeing disk size measured from
the guide star images, together with an estimated integration efficiency
based on the known brightness of and measured flux from the guide stars.
The observers can then track and compensate for focus changes and
dynamically adjust exposure times to match conditions, also using
information from the infrared sky camera.

We have easy-to-use tools incorporated into the guider software for
finding best telescope focus by sweeping through a range of secondary
mirror positions and plotting the resulting guide star image sizes at each
position. A focus sweep typically takes a few minutes.  On nights with
stable ambient temperatures, this may only be needed at the start of
night, and the observers will then use experience, helped by a simple
model of the thermal characteristics of the telescope, for subsequent
focus adjustments.  However, on some nights the temperature varies enough
to require more frequent focus sweeps with a consequent loss of observing
efficiency.  In efforts to find faster procedures, we have conducted two
sets of experiments to date.  The first was a differential focusing scheme
which tried to make use of the residual error between the plug plate
profiles and the true telescope focal surface to determine the direction
of best focus.  The second attempted a Hartmann scheme, closing half of
the flat-field petals at a time to produce a lateral shift in image
positions correlated with the direction of best focus.  In both cases the
results so far have been inconclusive.

Our aim when observing, as described in the next section,
is to achieve just sufficient signal-to-noise ratio to meet survey
requirements.  We take at least three target exposures to allow automatic
cosmic ray discrimination in the data reduction pipeline.  Under the best
conditions, these can be a short as 600~sec, rising to perhaps $5 \times
1500$~sec in cloud or very poor seeing.

For much of the survey through the end of 2003, observations of each plate
included a 4~min "smeared" exposure of the target that synthesised a
larger aperture intended for spectrophotometric calibration; these have
now been discontinued.\cite{abazajianetal2004}  Although the decision to
not apply this calibration was based on spectrophotometric
considerations,\cite{tremontietal2004} the decision to cease obtaining
smears clearly influenced observing efficiency.

\subsection{Real-time quality assurance}
\label{sec:qa}

Data quality assurance must start at the telescope, or much time can be
wasted either taking too much, too little, or simply bad data.

Our first aim is to ensure just enough photons are gathered for each
target.  Too few will not achieve the S/N required for the survey's
science goals, while too many wastes time exceeding those requirements.  A
second aim is that the observations are just sufficiently calibrated.  We
obtain at least one set of flat-field and arc lamp exposures for each
plate to trace fibre apertures in the images and apply a wavelength
calibration, but if there is too much flexure in the instruments during
target exposures, more sets may be required.  However, we do not want to
waste time taking extra calibration exposures if they are not required.

To measure S/N and flexure and to check for saturation and many other
problems in the data, we run a cut-down version of the full SDSS
spectrographic reduction pipeline, known as SoS ("Son of Spectro"), on a
fast dual-processor computer.  SoS operation is event-driven,
automatically giving results $\sim 2$ minutes after the end of an
exposure. It reports via web pages available immediately to observers and
off-site collaboration, giving clear warnings about flexure, stray light,
sky-line saturation, very bright objects and many more problems. It also
provides a table showing the summed S/N from all the good-quality science
exposures for a plate at canonical fibre magnitudes, and plots showing the
S/N distribution by magnitude and fibre position.  These help the
observers make informed decisions on both achieving the quality targets
and identifying and correcting problems.  Problems are identified as soon
as possible, which can result in big time savings.  For example, it is not
practical to combine science exposures from different pluggings of a
plate, so if an exposure must be discarded we need to know before we can
declare the plate finished and have it replaced with a new plate.
Obtaining a replacement science exposure on the same night also saves
unnecessary calibration exposures that would be required should
observations be carried over to extra nights.

\section{Flexible observing plans}
\label{sec:obsplan}

We switch between imaging and spectroscopy as observing conditions
dictate: imaging when the sky is moonless, photometric and seeing is $<
2$~arcsec FWHM, otherwise spectroscopy.  However, within those criteria,
we usually have a wide choice of the exact science targets to observe.
For spectroscopy, we now hold a queue of $\sim 350$-–$400$ plates at APO,
of which typically $\sim 100$ are observable during a given lunation.
Plates drilled for other seasons are held off site.

A plate database lets the observers choose which plates are to be plugged
with fibres ready for the next night so we can observe at any time during
the night.  The database displays information that makes planning around
current conditions easy.  The principal information shown is the position
on the sky of the target field for each plate and its observable time
range. This time range is determined by the airmass limits over which the
target image positions, which move because of changes in differential
refraction, remain acceptably close to the fibre positions on the
telescope focal surface.\cite{newman2002}  It hence varies from plate to
plate.  As there is a trade off between observed hour angle and the
observable time range, plates are drilled with fibre positions centred for
specific hour angles with the aim of maintaining a choice of plates
available for each hour of the night.

The plate database also logs and reports any S/N acquired for each plate
on preceding nights, and includes a handy ephemeris for the night giving
times of twilight and Moon rise and set and the degree of illumination
from the Moon at each plate's location on the sky.

When observations for a plate are completed, instructions are sent to the
plugging crew to replace the plate on that cartridge.  The observers
choose which new plates to mount, again based on information from the
plate database, so as to maintain the ability to make spectrographic
observations throughout the following night.

\section{Dedicated people}
\label{sec:people}

The term \emph{dedicated} is used here in two senses: first to mean that
most APO staff who work on the SDSS systems work \emph{only} on SDSS
systems,\footnote{Several of the SDSS engineering staff at APO may work on
other site systems if time permits.} and secondly, to say that the SDSS
staff at APO are \emph{enthusiastic\/} about the science that is coming
out of their work. In particular, all of the SDSS Observers are career
astronomers, most with doctorate-level education and experience, or with
strong backgrounds in observational and engineering practice, and all with
corresponding research interests.

Maintaining scientific motivation for the staff and hence, it is hoped,
employing them for long periods, has been important because the learning
curve for the SDSS systems is steep and long.  Experience has shown that
new observers take $\sim9$--12 months to get fully up to speed on all
systems.  This approach has proven successful: no SDSS staff have left APO
since 2000.

Among the observers, we have also put effort into developing individual
areas of expertise.  We recognise $\sim 70$ constituent systems, covering
hardware, software, procedures, training, observing plans, data reduction
and so on. We have split these up so that each observer concentrates on a
particular set. Each observer has thus taken responsibility for
understanding their set of systems, managing the recording and approval of
changes to those systems via the GNATS problem
reporting database (see next section), 
attending development meetings, writing procedures, and so on.  This has
proven to be a very successful division of effort, allowing each observer
to specialise on areas of interest and also build trust that the other
observers have taken similar charge of their own areas.

\section{Successes}
\label{sec:successes}

\subsection{Improved reliability}
\label{sec:reliable}

Although not quantified here, all of the SDSS observing software and
hardware systems at APO are now considerably more reliable than at the
start of the survey, due to increases in the mean time between failures
(MTBF) and decreases in mean time to repair (MTTR).  The improvement in
MTBF has come both from steadily ironing out bugs in the software and a
strong engineering programme of improving telescope and instrument
hardware and electronics.  The decreases in MTTR are at least partly due
to decreased workload on the staff responsible for the work, so that the
backlogs of outstanding work are shorter, but much project management
effort has also gone into tracking and managing the tasks involved.  All
software and documentation has been managed using the CVS version
management package, which has allowed close control of collaborative
development, testing and migration.

The entire survey maintains a database (known as GNATS) to track all
problems and change requests affecting all aspects of the SDSS systems
from construction through observing to data reduction pipelines and final
data distribution. To date, the database contains more than 6000 entries.
It has been very noticeable that the number of reported problems for the
observing systems has diminished greatly in the last year, as has the
frequency with which software version numbers have incremented.  Most new
problems now are associated with down-stream systems, indicating the
project has reached a certain level of maturity.

There is no doubt that the effort involved in careful management of change
has paid off.  It is doubtful that such a large collaborative effort could
be successful without it.

\subsection{Improved efficiency}

As with all repeated tasks, practice makes perfect, and some of the
observational efficiency gains since the start of the survey have come
simply from learning how to do things properly.  However, we have also
each worked hard on our particular areas of expertise to, for example,
remove redundant tasks and parallelise many observational steps.

As an example, the time from finishing the last exposure on one
spectrographic plate to acquiring the field for the next has been reduced
from a two-person $\sim 10$~min operation to a one-person $\sim 4$~min
operation by such changes as starting the slew to instrument change
position as soon as the exposure starts to read out from the CCDs, and by
orchestrating the redesign of the instrument latch controls to reduce the
distance the operator has to move to disengage and re-engage the
cartridges and to provide visual feedback on latch states.  These are but
two small examples of a continuing programme of improvements to procedures
and hardware that have greatly reduced the overheads associated with
actually gathering photons from the targets.

\subsection{Consistent spectral quality}

To date, the SDSS has obtained $\sim 6\times10^{5}$ spectra, of which more
than $3.6\times 10^5$ have been publicly released.  Just $\sim 0.5$ per
cent had insufficient S/N for secure spectral classification and radial
velocity.\cite{abazajianetal2004}  We consider this to be the greatest
measure of our success: the SDSS is truly a factory producing high-quality
spectra on an industrial scale.

\subsection{The record nights}

On each of the nights of MJD 52669 (2003 January 29/30) and MJD 53084
(2004 March 19/20), APO obtained 5,760 spectra using the SDSS systems -–
the most possible in one night with the SDSS equipment.  On those
record-setting nights, we were fortunate in that all the required
conditions came together: photometric sky, no moon, and seeing just too
bad to warrant survey imaging. Although these are the only nights so far
where we have achieved a ``max'', using all 9 available fibre-optic
plug-plate cartridges, we have observed 8 plates (5,120 spectra) on
several nights and 7 plates (4480 spectra) on many nights.

The candidates observed on MJD 52669 came from the SDSS programmes (which
includes the main\cite{strausetal2002} and luminous-red galaxy
samples,\cite{eisensteinetal2001} the quasar
sample,\cite{richardsetal2002} and other stellar and serendipitous samples
generated by the SDSS collaboration), plus two plates designed for
associated programmes exploring photometric redshifts and stellar
kinematics. On MJD 53084, all the plates were from the SDSS programmes.
The different classes of objects observed on each record night are listed
in table \ref{tab:candidates}.

\begin{table}[h]
  \centering
  \caption{Candidates observed on the record-setting nights. }
  \label{tab:candidates}
\begin{tabular}{ccccccccc}
    \hline \hline
    MJD & Plate & Programme & Galaxies & Quasars & Stars & Calibration &
    Sky& Others \\
    \hline
 52669 &  811 & Photo-z & 575 & 0 & 0 & 17  & 48  & 0 \\
        & 848  & SDSS    & 486 & 64  & 20  & 18  & 32  & 20 \\
        & 876 & SDSS    & 383 & 92  & 45  & 16  & 32  & 72 \\
        & 1130 & Kinematics & 0 & 0  & 592 & 16  & 32  & 0 \\
        & 1159    & SDSS    & 496 & 67  & 17  & 19  & 32  & 9 \\
        & 1163    & SDSS    & 516 & 73  & 1   & 17  & 32  & 1 \\
        & 1198    & SDSS    & 465 & 88  & 19  & 16  & 32  & 20 \\
        & 1203    & SDSS    & 357 & 121 & 57  & 16  & 32  & 57 \\
        & 1204    & SDSS    & 372 & 94  & 48  & 19  & 32  & 75 \\
        \cline{4-9}
        &         & Total   & 3650    & 599 & 799 & 154 & 304 & 254 \\

  \hline
 53084  & 1348    & SDSS    & 510 & 79  & 2   & 17  & 32  & 0 \\
        & 1368    & SDSS    & 490 & 93  & 3   & 16  & 32  & 6 \\
        & 1375    & SDSS    & 513 & 75  & 0   & 17  & 32  & 3 \\
        & 1379    & SDSS    & 494 & 96  & 1   & 16  & 32  & 1 \\
        & 1380    & SDSS    & 436 & 86  & 43  & 16  & 32  & 27 \\
        & 1440    & SDSS    & 472 & 70  & 20  & 17  & 32  & 29 \\
        & 1453    & SDSS    & 476 & 108 & 1   & 16  & 32  & 7 \\
        & 1619    & SDSS    & 520 & 69  & 0   & 17  & 32  & 2 \\
        & 1758    & SDSS    & 346 & 94  & 57  & 16  & 32  & 95 \\
        \cline{4-9}
        &         & Total      & 4257    & 770 & 127 & 148 & 288 & 170 \\
  \hline \hline
\end{tabular}
\end{table}

Of course, sky conditions, moon phase and variations in the length of the
night with season have a big influence on the observing rate we can
achieve, making it difficult to quote meaningful average rates.  However,
in typical conditions that are not good enough for imaging, we estimate
our median spectrographic observing rate at 6 plates (3840 spectra) per
night.


\section{Lessons learned}
\label{sec:lessons}

Probably the hardest section to write in an article such as this is the
admission of where we went wrong and the lessons that (we hope!) were
learned.  Of course, at a low level, many of the day-to-day problems
encountered in SDSS operations are very specific to this survey and hence
of minimal interest to others.  However, many higher-level issues have
been recognised since the survey started, and just a few of these with
impact on observing strategy and tactics will be mentioned here.

\subsection{Doing two things at once is difficult}

\ldots{Especially} when one has to precede the other.  The SDSS was
designed from the outset with the aim of gathering imaging and
spectrographic data on one telescope, with the spectrographic targets
being selected from the imaging data.  One of the most potentially
difficult scheduling problems at the start of a combined survey such as
this is the very necessity of obtaining imaging data before targets for
spectroscopy could be chosen. This was compounded by the science goals
that required observations throughout the year and the stringent observing
conditions required for achieving the imaging quality targets.

In fact, as it has turned out, we have very comfortably gained ground, so
that at the time of writing, the imaging survey is considerably ahead of
the spectral survey, so that we are no longer in any danger of running out
of spectral targets to observe.  The choice of APO as the site for the
SDSS telescopes has been justified not least by achieving the balance of
observing conditions required.

\subsection{Eventually, you have to (understand how to) select targets}

Another serious problem associated with achieving the survey homogeneity
objectives is that the spectrographic targeting
algorithms\cite{strausetal2002,eisensteinetal2001,richardsetal2002} were
all largely untested on the sky, and consequently underwent considerable
evolution from the start of the survey.  This has been a particular
concern for the quasar targeting, where the loci of galactic stars and
quasars cross for quasars at $z\simeq3$ in the 4-colour space from which
targets were selected.  This caused considerable difficulties in achieving
the desired redshift selection function, and the targeting algorithm was
not frozen until more than 16000 quasars ($\sim16$ per cent of the
expected survey total) had been observed.\cite{schneideretal2003}  One
penalty of this is that the homogeneous quasar data form a smaller subset
of the entire data set than would be desired.

\subsection{When you paint a floor, finish near an exit}

The aims of the SDSS are to image one quarter of the entire sky, then
obtain spectra of $\sim 1.1\times10^{6}$ objects from the same
area.\cite{yorketal2000}  Being a telescope based on Earth away from the
poles, the areas of sky actually accessible for observation varies with
season. This leaves us with the natural problem of ``painting ourselves
into corners'', whereby we have essentially finished large areas of the
sky for particular times of year.  This problem naturally gets worse as we
approach the completion of the survey.

There is therefore a clear need for follow-on programmes to fill the gaps
in the schedule. In this the SDSS is unusual, the entire system having
been built for the single purpose of conducting the survey, rather than as
a general purpose system.  Nevertheless, the collaboration seems to have
had little problem finding interesting things to do with the ``spare''
time that has arisen during the course of the survey, although new
mechanisms had to be grown to manage the allocation of time and resources
to these additional programmes.  As the end of the main survey comes into
sight, however, the problem will only get worse, and several major
proposals for extended follow-on programmes are in progress.  The lesson
learned here for special-purpose systems on the SDSS model is the need to
pay very great attention to the lead times between ideas and funding if
operational and staffing continuity is to be achieved.  In particular, the
risks of losing highly skilled and experienced staff because of a gap in
funding are immense, as once they have gone, any follow-on programmes
would be essentially starting from scratch in learning how to do things.

\subsection{Software will be more complicated than you imagined}

The software required to operate the entire SDSS is, as might be expected,
large and complex.  The code directly associated just with operating the
instruments consists of $\sim10^{5}$ lines of source code, not counting
common subroutines shared with other operational systems.  The version
number of this code is v3\_137\_0 which alone tells a long story, and that
is but one of the 18 distinct software systems that are used during any
night of operations.

Although all the software is now relatively stable (in the sense that it
does most of what is required and we know how to work around most of its
deficiencies!), it took a long time to reach this state, and as with many
one-off software development projects, ended up being far more complex
than was at first envisaged.  The first author of this article is not
alone in holding the opinion that a faster and better job could have been
done if the tasks of constructing the software had been managed by
professional software engineers instead of professional astronomers,
although, clearly, extensive astronomical insight was critical to defining
systems requirements.

\subsection{You will want to run the machines longer than you imagined}

Managing any special-purpose system into a long term asset requires a
strategy for evolution and maintenance of the hardware.  In the case of
the SDSS, there are several hardware systems that represent single-point
failure threats to survey operations. In particular, the imager is the
single most valuable asset, and a great deal of effort has gone into
protecting it. Nevertheless, we have already had several episodes where it
appeared for some time that we might lose one or another of the 54 CCDs it
incorporates.  Spares are limited and compatible replacement hardware
would be difficulty to obtain, so observational strategies for softening
the effects of failure are necessary. Similarly, any problems with the
complex optics and cameras in the spectrographs or any of the associated
data acquisition computers would be, if not show stoppers, at the very
least the cause of major slow-downs in the survey.

Although it would be nice to declare that everything is and will remain
copacetic, everyone is very much aware of all of these potential sources
of failure, and plans are in hand to provide short-term and long-term
solutions. But we are pleased to report that none of these problems has,
as yet, stopped us from continuing to build the largest extragalactic
redshift survey so far.


\acknowledgments     

Funding for the creation and distribution of the SDSS Archive has been
provided by the Alfred P. Sloan Foundation, the Participating
Institutions, the National Aeronautics and Space Administration, the
National Science Foundation, the U.S. Department of Energy, the Japanese
Monbukagakusho, and the Max Planck Society. The SDSS Web site is {\tt
http://www.sdss.org/}.

The SDSS is managed by the Astrophysical Research Consortium (ARC) for the
Participating Institutions. The Participating Institutions are The
University of Chicago, Fermilab, the Institute for Advanced Study, the
Japan Participation Group, The Johns Hopkins University, Los Alamos
National Laboratory, the Max-Planck-Institute for Astronomy (MPIA), the
Max-Planck-Institute for Astrophysics (MPA), New Mexico State University,
University of Pittsburgh, Princeton University, the United States Naval
Observatory, and the University of Washington.


\bibliography{ReferencesDB}   
\bibliographystyle{spiebib}   

\end{document}